\documentstyle[sprocl,epsfig]{article}

\def\be{\begin{equation}}
\def\ee{\end{equation}}
\def\bea{\begin{eqnarray}}
\def\eea{\end{eqnarray}}

\begin{document}
\title{THRESHOLD RESUMMATION OF SOFT GLUONS IN HADRONIC REACTIONS -- 
AN INTRODUCTION\footnote
{Invited paper to be published in the Proceedings of the Symposium on 
QCD Corrections and New Physics, Hiroshima, October 27 - 29, 1997.  
Argonne report ANL-HEP-CP-98-03.}}

\author{EDMOND L. BERGER}
\address{High Energy Physics Division
\\Argonne National Laboratory, 
\\Argonne, Illinois 60439, USA
\\E-mail: elb@hep.anl.gov}

\maketitle

\abstracts{
I discuss the motivation for resummation of the effects of initial-state
soft gluon radiation, to all orders in the strong coupling strength, for 
processes in which the near-threshold region in the partonic subenergy is 
important.  I summarize the method of ``perturbative resummation" and its 
application to the calculation of the total cross section for top 
quark production at hadron colliders.  Comments are included on the differences 
between the treatment of subleading logarithmic terms in this method and in 
other approaches.}

\section{Introduction and Motivation}

In inclusive hadron interactions at collider energies, 
$t\bar{t}$ pair production proceeds through partonic hard-scattering 
processes involving initial-state light quarks $q$ and gluons $g$.  In 
lowest-order perturbative quantum chromodynamics (QCD), at 
${\cal O}(\alpha_s^2)$,  the two partonic subprocesses 
are $q + \bar{q} \rightarrow t + \bar{t}$ and $g + g \rightarrow t + \bar{t}$.  
Calculations of the cross section through next-to-leading order, 
${\cal O}(\alpha_s^3)$, involve gluonic radiative corrections to these 
lowest-order subprocesses as well as contributions from the $q + g$ initial 
state~\cite{dawson}.  In this paper, I describe calculations that go 
beyond fixed-order perturbation theory through resummation of the effects of 
gluon radiation~\cite{laeneno,previous,catani} to all orders in 
the strong coupling strength $\alpha_s$.  

The physical cross section is obtained through the factorization theorem
\begin{equation}
\sigma_{ij}(S,m)=
{4m^2\over S}\int_0^{{S\over 4m^2}-1}d\eta
\Phi_{ij}(\eta,\mu) \hat\sigma_{ij}(\eta,m,\mu) .
\label{feleven}
\end{equation}
%\begin{eqnarray}
%&\sigma_{ij}(S,m)=
%{4m^2\over S}\int_0^{{S\over 4m^2}-1}d\eta&              \nonumber \\
%& \times \Phi_{ij}\biggl[{4m^2\over S}(1+\eta),\mu^2\biggr]
%\hat\sigma_{ij}(\eta,m^2,\mu^2)& .
%\label{feleven}
%\end{eqnarray}
%
The square of the total hadronic center-of-mass energy is $S$, the square of 
the partonic center-of-mass energy is $s$, $m$ denotes the top mass, $\mu$ is 
the usual factorization and renormalization scale, and $\Phi_{ij}(\eta,\mu)$ is 
the parton flux.  
The variable $\eta={s \over 4m^2} - 1$ measures the distance from the 
partonic threshold.  The indices $ij\in\{q\bar{q},gg\}$ denote the initial 
parton channel.  The partonic cross section 
$\hat\sigma_{ij}(\eta,m,\mu)$ is obtained either from fixed-order QCD
calculations~\cite{dawson}, or, as described here, from calculations
that include of resummation~\cite{laeneno,previous,catani} to all 
orders in $\alpha_s$.  I use the notation 
$\alpha \equiv \alpha(\mu=m) \equiv \alpha_s(m)/\pi$.  The total physical 
cross section is obtained after incoherent addition of the contributions from 
the $q\bar{q}$ and $gg$ production channels.  

Comparison of the partonic cross section at next-to-leading order with its 
lowest-order value reveals that the ratio becomes very large in the 
near-threshold region.  Indeed, as $\eta \rightarrow 0$, the ``$K$-factor" at 
the partonic level $\hat K(\eta)$ grows in proportion to $\alpha \ln^2(\eta)$. 
The very large mass of the top quark, and the correspondingly small value 
of $\alpha$ notwithstanding, the large ratio 
$\hat K(\eta)$ makes it evident that the next-to-leading order result does not 
necessarily provide a reliable quantitative prediction of the top quark 
production cross section at the energy of the Tevatron collider.  Analogous 
examples include the production of hadronic jets that carry large values of 
transverse momentum, the production of pairs of supersymmetric particles 
with large mass, and the pair-production of a fourth-generation quark, such as 
the postulated $b'$.  

\section{Gluon Radiation and Resummation}

The origin of the large threshold enhancement may be traced to initial-state
gluonic radiative corrections to the lowest-order channels.  
I remark that I am describing the calculation of the inclusive total cross 
section for the production of a top quark-antiquark pair, i.e., the total 
cross section for $t + \bar{t} + \rm anything$.  The partonic subenergy 
threshold in question is the threshold for $t + \bar{t} +$ any number of 
gluons.  This coincides with the threshold in the invariant mass of the 
$t + \bar{t}$ system for the lowest order subprocesses only. 

For $i + j \rightarrow t + \bar{t} + g$, the variable $z$ is defined through 
the invariant
$(1-z) = {2k \cdot p_t \over m^2}$, where $k$ and $p_t$ are the four-vector 
momenta of the gluon and top quark.  In the limit that 
$z \rightarrow 1$, the radiated gluon carries zero momentum.  After cancellation
of soft singularities and factorization of collinear singularities in 
${\cal O}(\alpha_s^3)$, there is a left-over integrable large logarithmic 
contribution to the partonic cross section associated with initial-state gluon 
radiation.  This contribution is often expressed in terms of ``plus" 
distributions.  In ${\cal O}(\alpha_s^3)$, it is proportional to 
$\alpha^3 \ln^2(1-z)$.  When integrated over the near-threshold 
region $1 \ge z \ge 0$, it provides an excellent approximation to the full 
next-to-leading order physical cross section as a function of the top 
mass~\cite{previous}.  

Although a fixed-order ${\cal O}(\alpha^4)$ calculation of 
$t\bar{t}$ pair production does not exist, universality of the form of 
initial-state soft gluon radiation may be invoked, and the leading logarithmic 
structure at  ${\cal O}(\alpha^4)$ may be appropriated from the 
next-to-next-to-leading order calculations of massive lepton-pair production 
($l\bar{l}$), the Drell-Yan process.  
In the near-threshold region, the hard kernel becomes
\begin{equation}
{\cal H}^{(0+1+2)}_{ij}(z,\alpha) \simeq 1+2\alpha C_{ij} \ln^2 (1-z)
+ \alpha^2\biggl[2C^2_{ij} \ln^4 (1-z) - {4\over 3} C_{ij} b_2
\ln^3 (1-z)\biggr] .
\label{btwo}
\end{equation}
The coefficient $b_2=(11C_A-2n_f)/12$; the number of flavors $n_f=5$; 
$C_{q\bar{q}}=C_F=4/3$; and $C_{gg}=C_A=3$.  
The leading logarithmic contributions in each order of perturbation 
theory are all positive in overall sign so that the leading 
logarithm threshold enhancement keeps building in magnitude at each fixed 
order of perturbation theory.

The goal of gluon resummation is to sum the series in $\alpha^n \ln^{2n}(1-z)$ 
to all orders in $\alpha$ in order to obtain a more trustworthy prediction.
This procedure has been studied extensively for the Drell-Yan process, and good 
agreement with data is achieved.  In essentially all resummation procedures, 
the large logarithmic contributions are exponentiated into a function of the 
QCD running coupling strength, itself evaluated at a variable momentum scale 
that is a measure of the radiated gluon momentum. 
The set of purely leading monomials $\alpha^n\ln^{2n}(1-z)$ exponentiates 
directly, with $\alpha$ evaluated at a fixed large scale $\mu = m$, as may 
be appreciated from a glance at Eq.~(\ref{btwo}). This simple result does not 
mean that a theory of resummation is redundant, even if only leading 
logarithms are to be resummed.  Indeed, straightforward use of the exponential 
of $\alpha 2 C_{ij}\ln^2(1-z)$ would lead to an exponentially divergent 
integral (and therefore cross section) since the coefficient of the logarithm 
is positive.  The naive approach fails, and more sophisticated resummation 
approaches must be employed.

Different methods of resummation differ in theoretically and phenomenologically
important respects.  Formally, if not explicitly in some approaches, an 
integral over the radiated gluon momentum $z$ must be done over regions in 
which $z \rightarrow 1$.  Therefore, one significant distinction among methods 
has to do with how the inevitable ``non-perturbative" region is handled.  
%In the approach of Laenen, Smith, and van Neerven (LSvN)~\cite{laeneno}, an 
%undetermined infrared  cutoff (IRC) $\mu_o$ is introduced, 
%with $\Lambda_{QCD} \leq \mu_o \leq m$.  
%The presence of an extra scale spoils the renormalization 
%group properties of the overall expression.  The unfortunate dependence of the 
%resummed cross section on this undetermined cutoff is important numerically 
%since it appears in an exponent~\cite{laeneno}.  

The method of resummation employed in my work with Harry 
Contopanagos~\cite{previous} is based on a perturbative truncation of 
principal-value (PV) resummation~\cite{stermano}.  
This approach has an important technical advantage in that it 
does not depend on arbitrary infrared cutoffs.  Because extra scales are 
absent, the method permits an evaluation of its regime of
applicability, i.e., the region of the gluon radiation phase
space where leading-logarithm resummation should be 
valid.  We work in the $\overline{\mbox{MS}}$ factorization scheme.

Factorization and evolution lead directly to exponentiation of the set of 
large threshold logarithms in moment ($n$) space in terms of an exponent 
$E^{PV}$.  The function $E^{PV}$ is finite, and 
$\lim_{n\rightarrow\infty}E^{PV}(n,m^2)=-\infty$.  Therefore, 
the corresponding partonic cross section is finite as $z\rightarrow 1
\ (n\rightarrow +\infty)$.

The function $E^{PV}$ includes both perturbative and non-perturbative content.  
The non-perturbative content is not a prediction of perturbative QCD. 
Contopanagos and I choose to use the exponent only in the interval in moment 
space in which the perturbative content dominates.  We derive a 
perturbative asymptotic representation of $E(x,\alpha(m))$ that is 
valid in the moment-space interval
\begin{equation}
1<x\equiv \ln n< t\equiv {1\over 2\alpha b_2}.
\label{tseven}
\end{equation}
The interval in 
Eq.~(\ref{tseven}) agrees with the intuitive definition of the perturbative 
region in which inverse-power contributions are unimportant: 
$\Lambda_{QCD} \over{(1-z)m}$ $\le 1$.

The perturbative asymptotic representation is
\begin{equation}
E_{ij}(x,\alpha)\simeq E_{ij}(x,\alpha,N(t))=
2C_{ij}\sum_{\rho=1}^{N(t)+1}\alpha^\rho
\sum_{j=0}^{\rho+1}s_{j,\rho}x^j\ .
\label{teight}
\end{equation}
Here
\begin{equation}
s_{j,\rho}=-b_2^{\rho-1}(-1)^{\rho+j}2^\rho c_{\rho+1-j}(\rho-1)!/j!\ ;
\label{tnine}
\end{equation}
and $\Gamma(1+z)=\sum_{k=0}^\infty c_k z^k$, where $\Gamma$ is the Euler gamma 
function.  
The number of perturbative terms $N(t)$ in Eq.~(\ref{teight}) is
obtained~\cite{previous} by optimizing the asymptotic approximation
$\bigg|E(x,\alpha)-E(x,\alpha,N(t))\bigg|={\rm minimum}$. 
Optimization works perfectly, with $N(t)=6$ at $m = 175$ GeV.   
As long as $n$ is in the interval of Eq.~(\ref{tseven}),
all the members of the family in $n$ are optimized 
at the same $N(t)$, showing that the optimum number of 
perturbative terms is a function of $t$, i.e., of $m$ only.

Resummation is completed in a finite number of steps.  When the running 
of the coupling strength $\alpha$ is included up to two loops, 
all monomials of the form $\alpha^k\ln^{k+1}n,\ \alpha^k\ln^kn$
are produced in the exponent of Eq.~(\ref{teight}).  We discard monomials
$\alpha^k\ln^kn$ in the exponent because of the restricted leading-logarithm
universality between $t\bar{t}$ production and massive lepton-pair 
production, the Drell-Yan process.

The moment-space exponent that we use is the truncation
\begin{equation}
E_{ij}(x,\alpha,N)=2C_{ij}\sum_{\rho=1}^{N(t)+1}\alpha^\rho s_\rho x^{\rho+1} ,
\label{tseventeen}
\end{equation}
with the coefficients
$s_\rho\equiv s_{\rho+1,\rho}=b_2^{\rho-1}2^\rho/\rho(\rho+1)$.  This
expression contains no factorially-growing (renormalon) terms. 
One can also derive the perturbative expressions,
Eqs.~(\ref{tseven}), (\ref{teight}), and (\ref{tnine}),  without the 
principal-value prescription, although with less certitude~\cite{previous}.

After inversion of the Mellin transform from moment space to the physically 
relevant momentum space, the resummed hard kernel takes the form 
\begin{equation} 
{\cal H}^{R}_{ij}(z,\alpha)=\int_0^{\ln({1\over 1-z})}
dx{\rm e}^{E_{ij}(x,\alpha)}
\sum_{j=0}^\infty Q_j(x,\alpha)
\ .\label{bafour}
\end{equation}
The leading large threshold corrections are contained in the exponent 
$E_{ij}(x,\alpha)$, a calculable polynomial in $x$.  The functions 
$\{Q_j(x,\alpha)\}$ arise from the analytical inversion of the Mellin 
transform from moment space to momentum space. 
These functions are expressed in 
terms of successive derivatives of $E$. Each 
$Q_j$ contains $j$ more powers of $\alpha$ than of $x$ so that 
Eq.~(\ref{bafour}) embodies a natural power-counting of threshold 
logarithms.  However, only the {\it leading} 
threshold corrections are universal. Final-state gluon radiation as well as
initial-state/final-state interference effects produce subleading logarithmic 
contributions that differ for processes with different final states.  
Accordingly, there is no physical basis for accepting the validity of the 
particular subleading terms that appear in Eq.~(\ref{bafour}).  Among 
all $\{Q_j\}$ in Eq.~(\ref{bafour}), only the very leading one is universal, 
$Q_0$, and it is the only one we retain.  Hence, Eq.~(\ref{bafour}) can be 
integrated explicitly, and the resummed partonic cross sections become 
\begin{equation}
\hat{\sigma}_{ij}^{R;pert}(\eta,m)=
\int_{z_{min}}^{z_{max}}dz
{\rm e}^{E_{ij}(\ln({1\over 1-z}),\alpha)}
\hat{\sigma}_{ij}'(\eta,m,z) .
\label{bthreep}
\end{equation}
%
%The leading large threshold corrections are contained in the exponent 
%$E_{ij}(x,\alpha)$, a calculable polynomial in $x$.  
The derivative
$\hat{\sigma}_{ij}'(\eta,m,z)=d(\hat{\sigma}_{ij}^{(0)}(\eta,m,z))/dz$,
and $\hat{\sigma}_{ij}^{(0)}$ is the lowest-order ${\cal O}(\alpha_s^2)$
partonic cross section expressed in terms of inelastic kinematic variables.
The lower limit of integration, $z_{min}$, is fixed by kinematics.  The upper 
limit, $z_{max} < 1$, well specified within the context of our 
calculation, is established by the condition of consistency of 
leading-logarithm resummation.  It is derived from the requirement that the 
value of all subleading contributions $Q_j, j\ge 1$ be negligible compared to 
the leading contribution $Q_0$.  The presence of $z_{max}$ guarantees that 
the integration over the soft-gluon 
momentum is carried out only over a range in which poorly specified 
non-universal subleading terms would not contribute significantly even if 
retained.  We cannot justify continuing the results of leading-logarithm 
resummation into the region $1 > z > z_{max}$.

To obtain the physical cross section, we insert the resummed expression 
Eq.~(\ref{bthreep}) into Eq.~(\ref{feleven}) and integrate over $\eta$.  
Perturbative resummation probes the threshold down to
$\eta\ge \eta_0 =(1-z_{max})/2 $.  Below this value, perturbation theory is 
not to be trusted.  For $m$ = 175 GeV, we determine that the perturbative 
regime is restricted to values of the subenergy greater than 1.22 GeV above the 
threshold ($2m$) in the $q{\bar q}$ channel and 8.64 GeV 
above threshold in the $gg$ channel. The difference reflects the larger
color factor in the $gg$ case.  The value 1.22 GeV is comparable to the decay 
width of the top quark, a natural definition of the perturbative boundary and 
by no means unphysically large.    

\section{Physical cross section}

Other than the top mass, the only undetermined scales are the QCD 
factorization and renormalization scales.  A common value $\mu$ is adopted for 
both.
\begin{figure}
%[b!]
%\vspace{1cm}
\centerline{\epsfig{file=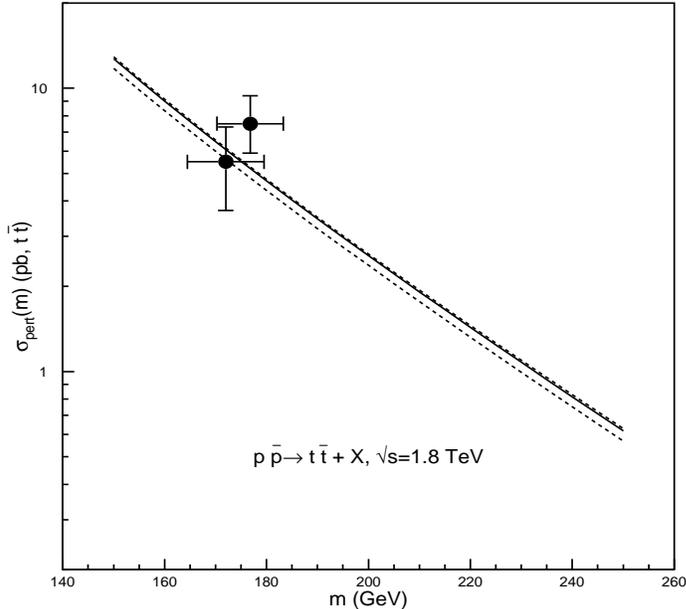,height=2.75in,width=2.75in}}
%\vspace{10pt}
\caption{Inclusive total cross section for top quark production.  The dashed 
curves show the upper and lower limits while the solid curve 
is our central prediction.  CDF and D0 data are shown.}
\label{fig1}
\end{figure}
In Fig.~1, our total cross section for $t\bar{t}$-production is shown  
as a function of top mass in $p \bar{p}$ collisions 
at $\sqrt{S}=1.8$ TeV.  The central value is obtained with the 
choice $\mu/m=1$, and the lower and upper limits are  the maximum 
and minimum of the cross section in the range $\mu/m\in\{0.5,2\}$.  
At $m =$ 175 GeV, the full width of this uncertainty band 
is about 10\%\ .  As is to be expected, less variation with $\mu$ is evident 
in the resummed cross section than in the next-to-leading order cross section.  
In estimating uncertainties, Contopanagos and I do not consider explicit 
variations of the non-perturbative boundary, expressed through $z_{max}$.  
For a fixed $m$ and $\mu$, $z_{max}$
is obtained by enforcing dominance of the universal leading logarithmic terms 
over the subleading ones. Therefore, $z_{max}$
is {\it derived} and is not a source of uncertainty.  At fixed $m$, the 
boundary necessarily varies as $\mu$ and thus $\alpha$ vary. 

Contopanagos and I calculate 
$\sigma^{t\bar{t}}(m=175\ {\rm GeV},\sqrt{S}=1.8\ {\rm TeV})=
5.52^{+0.07}_{-0.42}\ \rm{pb}$, in agreement with data~\cite{cdfdz}.  
This cross section is larger than the 
next-to-leading order value by about $9\%$.  The top quark cross section 
increases quickly with the energy of the $p \bar{p}$ collider.  We determine 
$\sigma^{t\bar{t}}(m=175\ {\rm GeV},\sqrt{S}=2\ {\rm TeV})=
7.56^{+0.10}_{-0.55}\ \rm {pb}$.  The central value rises to 22.4 pb at 
$\sqrt{S}=3\ {\rm TeV}$ and 46 pb at $\sqrt{S}=4\ {\rm TeV}$.

Extending our calculation to larger values of 
$m$ at ${\sqrt S}=1.8$ TeV, we find that resummation in the principal 
$q\bar{q}$ channel produces 
enhancements over the next-to-leading order cross section of $21\%$, $26\%$, 
and $34\%$, respectively, for $m =$ 500, 600, and 700 GeV.  The reason for the
increase of the enhancements with mass at fixed energy is that the threshold 
region becomes increasingly dominant.  Since the $q\bar{q}$ 
channel also dominates in the production of hadronic jets at very large values 
of transverse momenta, we suggest that on the order of $20\%$ of the excess
cross section reported by the CDF collaboration~\cite{cdfjets} may be 
accounted for by resummation.

\section{Other Methods of Resummation}
Two other groups have published calculations of the total cross section at 
$m=175\ {\rm GeV}$ and $\sqrt{S}=1.8\ {\rm TeV}$:
$\sigma^{t\bar t}$({\rm LSvN}~\cite{laeneno}) = $4.95^{+0.70}_{-0.40}$ pb; 
and  
$\sigma^{t\bar t}$({\rm CMNT}~\cite{catani}) = $4.75^{+0.63}_{-0.68}$ pb.  
From a numerical point of view, ours and theirs all agree 
within their estimates of theoretical uncertainty.  However, the resummation 
methods differ as do the methods for estimating uncertainties.  Both 
the central value and the band of uncertainty of the LSvN predictions are 
sensitive to their arbitrary infrared cutoffs.  To estimate theoretical 
uncertainty, Contopanagos and I 
use the standard $\mu$-variation, whereas LSvN obtain theirs 
primarily from variations of their cutoffs.  It is difficult to be certain 
of the central value and to evaluate theoretical uncertainties in a method 
that requires an undetermined infrared cutoff.  

The group of Catani, Mangano, Nason, and Trentadue (CMNT)~\cite{catani} 
calculate a central value of 
the resummed cross section (also with $\mu/m = 1$) that is less than 
$1\%$ above the exact next-to-leading order value.  
%There are similarities and differences between our approach 
%and the method of CMNT.  
Both they and we use the same universal leading-logarithm 
expression in moment space, but differences occur after the transformation to
momentum space.  The differences can 
be stated more explicitly if one examines the perturbative expansion of the
resummed hard kernel ${\cal H}^{R}_{ij}(z,\alpha)$.   
If, instead of 
restricting the resummation to the universal leading logarithms, one  
uses the full content of ${\cal H}^{R}_{ij}(z,\alpha)$, she or he would find 
an analytic expression that is equivalent to CMNT's numerical inversion, 
\begin{equation}
{\cal H}^{R}_{ij} \simeq 1+2\alpha C_{ij} 
\biggl[\ln^2 (1-z) + 2\gamma_E \ln (1-z)\biggr] + {\cal O}(\alpha^2).
\label{padovao}
\end{equation} 
In terms of this expansion, in our work we retain only the leading term 
$\ln^2 (1-z)$ at order $\alpha$, but both this term and the non-universal 
subleading term $ 2\gamma_E \ln (1-z)$ are retained by CMNT.  
If this subleading term is discarded in Eq.~(\ref{padovao}), the
residuals $\delta_{ij}/\sigma_{ij}^{NLO}$ defined by CMNT increase from 
$0.18\%$ to $1.3\%$ in the $q\bar{q}$ production channel and from $5.4\%$ to 
$20.2\%$ in the $gg$ channel.  After addition of the two 
channels, the total residual $\delta/\sigma^{NLO}$ grows from the negligible 
value of about $0.8\%$ to the value $3.5\%$.  While still smaller than 
the increase of $9\%$ that we obtain, the increase of $3.5\%$ vs. $0.8\%$ 
shows the substantial influence of the subleading logarithmic terms in the 
CMNT results.

Contopanagos and I 
judge that it is preferable to integrate over only the region of phase 
space in which the subleading term is suppressed numerically.  Our reasons 
include the fact that the subleading term is not universal, is not the same 
as the subleading term in the exact ${\cal O}(\alpha^3)$ calculation, and 
can be changed if one elects to keep non-leading terms in moment space.  The 
subleading term is negative and numerically very significant when it is  
integrated throughout phase space (i.e., into the region of $z$ above our 
$z_{max}$).  
%In the $q\bar{q}$ channel at $m=175$ GeV and ${\sqrt S}=1.8$ TeV, 
%its inclusion eliminates more than half of the contribution from the leading 
%term.  Although the goal is to resum the threshold contributions responsible 
%for the enhancement of the cross section at next-to-leading order, the method 
%of Ref.~\cite{catani} does not reproduce most of this enhancement.  The 
%influence of non-universal subleading terms is amplified at higher orders 
%where additional subleading structures occur in the approach of 
%Ref.~\cite{catani} with significant numerical coefficients proportional 
%to $\pi^2$, $\zeta(3)$, and so forth.  
In our view, the results of a 
{\it leading-logarithm} resummation should not rely on subleading structures 
in any significant manner.  The essence of our determination of the 
perturbative boundary $z_{max}$ is precisely that below $z_{max}$ subleading 
structures are also numerically subleading, whether or not these poorly 
substantiated subleading logarithms are included.
%Since the large threshold logarithms are mastered by resummation, the 
%theoretical reliability of the resummed result is greater than 
%that of a fixed-oder calculation.  Our analysis and the stability of our cross 
%section under variation of the hard scale $\mu$ provide confidence that our 
%perturbative resummation yields an accurate calculation of the inclusive top 
%quark cross section at Tevatron energies.
%

In the remainder of this section I offer a more systematic 
analysis~\cite{latest} of the role played in the CMNT approach by 
non-universal subleading 
logarithms and show in some detail how their method and results differ from 
ours.  I treat expansions of the resummed momentum-space kernel up to 
two loops. The corresponding
cross sections are integrable down to threshold, $z_{max} = 1$ and $\eta = 0$. 
However, the effects of the various classes of logarithms are 
pronounced if one continues the region of integration beyond our 
perturbative regime. 

In moment space, the exponent to two-loops is obtained from Eq.~(\ref{teight}):
\begin{equation}
E_{ij}^{[2]}(x,\alpha)=g \alpha(s_{2,1}x^2+s_{1,1}x+s_{0,1})
+g \alpha^2(s_{3,2}x^3+s_{2,2}x^2+s_{1,2}x+s_{0,2}) ,
\label{extraone}
\end{equation}
with $g=2C_{ij}$ and $x = \ln n$.  One can perform the analytical Mellin 
inversion directly,  beginning with Eq.~(\ref{extraone}). 
After a trivial integration, the results for the one- and 
two-loop hard kernels are 
\begin{equation}
{\cal H}^{(1)}=x_z^2\alpha\{g s_{2,1}\}
+x_z\alpha\{g(s_{1,1}+2c_1s_{2,1})\}\ ,
\label{extratwo}
\end{equation}
and
\begin{eqnarray} 
& &{\cal H}^{(2)}=x_z^4\alpha^2\{g^2s_{2,1}^2/2\}
+x_z^3\alpha^2\{gs_{3,2}+g^2(s_{2,1}s_{1,1}+
2c_1s_{2,1}^2)\}\nonumber \\
& &+x_z^2\alpha^2\{g(s_{2,2}+3c_1s_{3,2})+g^2(s_{1,1}^2/2+3c_1s_{1,1}s_{2,1}+
s_{2,1}s_{0,1}+  s_{2,1}^2[6c_2-\pi^2]\}\nonumber \\
& &+x_z\alpha^2\{g(s_{1,2}+2c_1s_{2,2}+s_{3,2}[6c_2-\pi^2])\nonumber \\
& &+g^2(s_{0,1}s_{1,1}
+2c_1s_{0,1}s_{2,1}+c_1s_{1,1}^2+s_{2,1}s_{1,1}[6c_2-\pi^2]\nonumber \\
& &+s_{2,1}^2[12c_3-2\pi^2c_1])\}\ .
\label{extrathree}
\end{eqnarray}
All the constants are defined in Eqs.~(\ref{teight}) and (\ref{tnine}).  
Equation~(\ref{extratwo}) includes a leading logarithmic term, 
$x_z^2\alpha$, as well as a next-to-leading term, $x_z\alpha$.  

The question to be addressed is whether it is justified and 
meaningful to retain 
all of the terms in Eqs.~(\ref{extratwo}) and (\ref{extrathree}) in the 
computation of the resummed cross section.  The issue has to do with what one 
intends by resummation of leading logarithms.  Contopanagos and I 
use the term {\it leading logarithm} resummation to denote the case in 
which the moment space exponent, Eq.~(\ref{extraone}), contains only the 
constants $E_{LL}=\{s_{\rho+1,\rho},0\}$.  This is also what is done in the 
CMNT method, and the exponent {\it in moment space} in 
their work is identical to that used for our predictions, 
Eq.~(\ref{tseventeen}).  However, in contrast to our expression in momentum 
space, Eq.~(\ref{bthreep}), the corresponding CMNT expression in momentum 
space includes the numerical equivalent of all 
terms in Eqs.~(\ref{extratwo}) and (\ref{extrathree}) that are proportional to 
$s_{\rho+1,\rho}$.

If expressed analytically, CMNT's corresponding ``LL" hard kernels are 
\begin{equation}
{\cal H}^{(1)}_{LL}=x_z^2\alpha g-x_z\alpha2g\gamma_{E} ,
\label{extrafour}
\end{equation}
and
\begin{eqnarray}
& &{\cal H}^{(2)}_{LL}=x_z^4\alpha^2g^2/2+x_z^3\alpha^2\{2gb_2/3-2\gamma_Eg^2\}
\nonumber \\
& &+x_z^2\alpha^2\{-2gb_2\gamma_E+g^2[3\gamma_E^2-\pi^2/2]\}\nonumber \\
& &+x_z\alpha^2\{2gb_2[3\gamma_E^2-\pi^2/2]/3+g^2[\gamma_E\pi^2-2\gamma^3-
4\zeta(3)]\}\ ,
\label{extrafive}
\end{eqnarray} 
where $\zeta(s)$ is the Riemann zeta function; $\zeta(3)=1.2020569$.
%The subleading terms in momentum space that CMNT retain in their ``LL" 
%resummation are non-universal and are multiplied by large negative constants. 
Evaluating the expressions numerically for the $q{\bar q}$ channel, one 
obtains~\cite{latest}  
\begin{equation}
{\cal H}^{(1)}_{LL}=x_z^2\alpha\times 2.66666-x_z\alpha\times 3.07848 ,
\label{extrasix}
\end{equation}
and
\begin{eqnarray}
& &{\cal H}^{(2)}_{LL}=x_z^4\alpha^2\times 3.55555
-x_z^3\alpha^2\times 4.80189 \nonumber \\
& &-x_z^2\alpha^2\times 33.88456-x_z\alpha^2\times9.82479\ .
\label{extraseven}
\end{eqnarray}
Apart from the leading monomials that are the same as those in our approach, 
Eqs.~(\ref{extrasix}) and (\ref{extraseven}) include a series of 
subleading terms, each of which has a significant negative coefficient. 
In practice, these subleading terms suppress the effects of resummation 
essentially completely.  One of the effects of this suppression is that the 
resummed partonic cross section is {\it smaller} than its next-to-leading 
order counterpart in the neighborhood of $\eta =$ 0.1, a region in which the 
next-to-leading order partonic cross section takes on its largest values.  This 
point is illustrated in Fig.~3 of CMNT's second paper~\cite{catani}.  

Although the specific set of subleading terms in 
Eqs.~(\ref{extrasix}) and (\ref{extraseven}) is generated in the inversion of 
the Mellin transform, a case can be made that the terms are accidental.  
%Our reasoning is based on an examination of the exact next-to-leading order 
%calculation of the cross section for heavy quark production and of similar 
%calculations of the Drell-Yan process up to two-loops.  
First, terms involving $\gamma_E$ do not appear in the exact next-to-leading 
order calculation of the hard part, since they are removed in the specification 
of the ${\overline{\rm MS}}$ factorization scheme.  Therefore, the term 
proportional to $\gamma_E$ in Eq.~(\ref{extrafour}) is suspect.  Second, if 
the specific value of the subleading logarithm is extracted from the full 
${\cal O}(\alpha^3)$ next-to-leading order calculation, one  
finds~\cite{latest} $x_z\alpha(2g - 41/6)$ instead 
of the term $-x_z\alpha2g\gamma_{E}$.  Instead of the numerical 
coefficient 3.07848 in 
Eq.~(\ref{extrasix}), one finds the smaller value 1.5 if the subleading 
logarithm of the exact ${\cal O}(\alpha^3)$ calculation is used.  Thus, not 
only is the ${\cal O}(\alpha)$ subleading term retained in the CMNT approach 
different from that of the exact calculation, it is 
numerically about twice as large.  Third, the results of a 
LL resummation should not rely on the subleading structures in any significant 
way.  However, in the CMNT approach, 
Eq.~(\ref{extrafour}), which is the one-loop projection 
of their resummed prediction, reproduces only 1/3 of the exact 
${\cal O}(\alpha^3)$ enhancement, the other 2/3 being cancelled by the 
second (non-universal) term of Eq.~(\ref{extrafour}).  Although the goal is 
%of Ref.~\cite{catani} fails an important consistency check: it 
to resum the threshold corrections responsible for the large enhancement of 
the cross section at next-to-leading order, the CMNT method does not reproduce 
most of this enhancement.

Addressing questions associated with the $\gamma_E$ terms, 
CMNT examine a type of NLL resummation in their second paper~\cite{catani}.  
In this NNL resummation, the 
%$E_{NLL}=
$\{s_{\rho+1,\rho},s_{\rho,\rho}\}$ terms are retained in the 
exponent, Eq.~(\ref{extraone}).  The corresponding hard kernels become 
\begin{equation}
{\cal H}^{(1)}_{NLL}=x_z^2\alpha g ,
\label{extraeight}
\end{equation}
and
\begin{equation}
{\cal H}^{(2)}_{NLL}=x_z^4\alpha^2g^2/2+x_z^3\alpha^22gb_2/3
-x_z^2\alpha^2g^2[\gamma_E^2+\pi^2/2]-x_z\alpha^2\{gb_2[2\gamma_E^2+\pi^2/3]
+g^24\zeta(3)\}\ .
\label{extranine}
\end{equation}
%
%Comparing Eqs.~(\ref{extraeight}) and (\ref{extranine}) with 
%Eq.~(\ref{btwo}), we observe that 
Equation~(\ref{extraeight}) is identical to the one-loop projection of our 
hard kernel.  
%As shown in 
%Ref.\cite{lgpaper}, it yields an excellent approximation to the 
%exact next-to-leading 
%order cross section.  
On the other hand, our two-loop projection contains only 
the first two terms of Eq.~(\ref{extranine}).   The term proportional to 
$x_z^3\alpha^2$ is present in our case, along with the leading term 
proportional to $x_z^4\alpha^2$, because it comes from the leading 
logarithms in the exponent $E(n)$, through two-loop running of the coupling 
strength.  In contrast to 
Eq.~(\ref{extrafive}), Eq.~(\ref{extranine}) relegates the influence 
of the ambiguous constant coefficients to lower powers of $x_z$ (but with 
larger negative coefficients).  In the amended scheme, the unphysical 
$\gamma_E$ terms are still present in the two-loop result, 
Eq.~(\ref{extranine}), along with $\pi^2$ and $\zeta(3)$ terms that may be 
expected but whose coefficients have no well defined physical origin.  
Recast in numerical form, Eqs.~(\ref{extraeight}) and (\ref{extranine}) 
become~\cite{latest}  
\begin{equation}
{\cal H}^{(1)}_{NLL}=x_z^2\alpha\times 2.66666 ,
\label{extraninep}
\end{equation}
and
\begin{equation}
{\cal H}^{(2)}_{NLL}=x_z^4\alpha^2\times 3.55555+x_z^3\alpha^2\times 3.40739
-x_z^2\alpha^2\times 37.46119-x_z\alpha^2\times 54.41253\ .
\label{extraten}
\end{equation}
There is a significant difference between the coefficients of 
all but the very leading power of $x_z$ in 
Eqs.~(\ref{extrasix}) and (\ref{extraseven}) with respect to those in  
Eqs.~(\ref{extraninep}) and (\ref{extraten}), and the 
numerical coefficients grow in magnitude as the power of $x_z$ decreases.  

Using their NLL amendment, CMNT find that the central value of their resummed 
cross section exceeds the next-to-leading order 
result by $3.5\%$ (both $q{\bar q}$ and $gg$ channels added). This increase 
is about 4 times larger than the central value of the increase obtained in 
their first method, closer to our increase of about $9\%$.   The reason for the 
significant change of the increase resides with the subleading structures, 
viz., in the differences between the LL version 
Eqs.~(\ref{extrasix}) and (\ref{extraseven}) and the NLL version 
Eqs.~(\ref{extraninep}) and (\ref{extraten}).  The subleading terms at 
two-loops cause a total suppression of the two-loop contribution (in fact, 
that contribution is negative), if one integrates all the way into what we 
call the non-perturbative regime.  This suppression explains why an 
enhancement of only $3.5\%$ is obtained in the amended method, rather than 
our $9\%$.

CMNT argue that retention of their subleading terms in momentum space is 
important for ``energy conservation".  By this statement, they mean that one 
begins the formulation of resummation with an expression in momentum space  
containing a delta function representing conservation of the fractional 
partonic momenta.  In moment space, this delta function subsequently 
unconvolves the resummation.   Therefore, when one inverts the Mellin transform
to return to momentum space, the full set of logarithms generated by this 
inversion are required by the original energy conservation.  This line of 
reasoning would be compelling {\it if the complete exponent $E(n)$ in moment 
space were known exactly}, i.e., if the resummation in moment space were exact 
in representing the cross section to all orders.  However, the exponent is 
truncated in all approaches, and knowledge of the logarithms it resums 
reliably is limited both in moment and in momentum space.  Hence, the set of 
logarithms produced by the Mellin inversion in momentum space should also be 
restricted.  In our approach energy conservation is obeyed in momentum space 
consistently with the class of logarithms resummed.  On the other hand, in the 
CMNT method, knowledge is presumed of all logarithms 
generated from the Mellin inversion, despite the fact that the truncation in 
moment space makes energy conservation a constraint restricted to the class of 
logarithms that is resummable, i.e., a constraint restricted by the truncation 
of the exponent $E(n)$.  The two approaches would be equivalent provided a 
constraint be in place on the effects of subleading logarithms.  This  
constraint is precisely our restriction $z_{max} < 1$, but no such 
constraint is furnished by CMNT.  For this reason their results 
change significantly if one set of the 
logarithms generated in momentum space is adopted as ``the set corresponding 
to energy conservation", and then compared with another set, produced by a 
different truncation of $E(n)$. 

%We have identified the terms responsible for the difference between our answer 
%for the resummed cross section and that of CMNT.  These differences reside 
%with subleading logarithms whose presence is not substantiated by physical 
%arguments.  
The essence of our determination of the perturbative regime, 
$z_{max} < 1$, is precisely that, in this regime, 
subleading structures are also {\it numerically subleading}, whether or not 
the classes of subleading logarithms coming from different truncation of 
the master formula for the resummed hard kernel are 
included.  The results presented in Fig.~11 of our 
second paper~\cite{previous}, show 
that if we alter our resummed hard kernel to account for subleading
structures but still stay within our perturbative regime, the resulting 
cross section is reduced by about $4\%$, within our band of perturbative 
uncertainty.

A criticism~\cite{catani} is that of putative ``spurious factorial
growth'' of our resummed cross section, above and beyond the infrared 
renormalons that are eliminated from our approach.  The issue, as 
demonstrated in Eq.~(29) of our second paper~\cite{previous}, can be addressed 
most easily if one substitutes any monomial appearing in 
Eq.~(\ref{extrathree}), symbolically 
$\alpha^mc(l,m)\ln^l x_z$, into Eq.~(\ref{bthreep}) and integrates over $z$:  
\begin{equation}
\alpha^mc(l,m)\int_{z_{min}}^1dz\ln^l x_z=\alpha^mc(l,m)(1-z_{min})l!
\sum_{j=0}^l\ln^j(1/(1-z_{min}))\ .
\label{uclaone}
\end{equation}
For the purposes of this demonstration we set $\hat \sigma'_{ij}=1$.  
The coefficients $c(l,m)$ can be read directly from 
Eq.~(\ref{extrathree}).  For the leading logarithmic terms, 
\begin{equation}
c(2m,m) \propto 1/m! ,
\label{uclaonep}
\end{equation} 
where this factorial comes directly from exponentiation.   After the 
integration over the entire $z$-range, the power of the logarithm in $x_z$ 
becomes a factorial multiplicative factor, $l!$. The presence of $l!$ 
follows directly from the existence of the powers of $\ln x_z$ 
that are present explicitly in the finite-order result in pQCD and is therefore 
inevitable.
%the former are spurious, so are the latter.  
If this exercise is 
repeated, but with the range of integration in Eq.~(\ref{uclaone}) constrained 
to our perturbative regime, one obtains the difference between the 
right-hand-side of Eq.~(\ref{uclaone}) and a similar expression containing 
$z_{max}$.  The result is numerically smaller, but both of the pieces are 
multiplied by $l!$.  

The factorial coefficient $l!$ is not the most important source of 
enhancement.  
For the leading logarithms at two-loop order, $l=2m=4$, and the overall 
combinatorial coefficient from Eqs.~(\ref{uclaone}) and (\ref{uclaonep}) is 
$(2m)!/m!=12$.  For comparison, at representative values of $\eta$ near 
threshold, $\eta=0.1$ and 0.01, the sum of logarithmic terms in 
Eq.~(\ref{uclaone}) provides factors of 16.1 and 314.3, respectively.  
Similarly, the (multiplicative) color factors at this order of perturbation 
theory are $(2C_{ij})^2=$ 7.1 and 36 for the $q\bar{q}$ and $gg$ channels, 
respectively.  All of these features are connected to the way 
threshold logarithmic contributions appear in finite-order pQCD and how they 
signal the presence of the non-perturbative regime.  Thus, 
preoccupation with the $l!$ factor seems misplaced~\cite{latest}.

%The phrase ``spurious factorial growth'' appears to rename 
%the logarithmic enhancements present in Eqs.~(\ref{extratwo}) and 
%(\ref{extrathree}), after the integral over $z$.  On the other hand, 
%according to our understanding, 
%The claim~\cite{catani} 
``Absence of factorial growth" is based on the use by CMNT 
of Eq.~(\ref{extraseven}) for their main predictions,
an expression that contains non-universal subleading logarithms, all with
significant negative coefficients.  Mathematically, factorial growth is present 
for each of the powers of the logarithm in Eq.~(\ref{uclaone}), since 
these monomials are linearly independent.  Absence of 
factorial growth based on a numerical cancellation between various classes 
of logarithms, most of them with physically unsubstantiated coefficients, 
appears to us to be an incorrect use of terminology.
% rather than a transparent
%expression of the mathematics.  
%From a purely phenomenological point of view, one cannot claim 
%that a $9\%$ increase of the top quark cross section at $m=175\ {\rm GeV}$ 
%and $\sqrt{S}=1.8\ {\rm TeV}$ reveals factorial growth but that an  
%$0.8\%$ increase does not.  
In the CMNT approach 
the effects of resummation are suppressed by a series of subleading logarithms 
with large negative coefficients.  If there is no physical basis for 
preference of Eqs.~(\ref{extrafour}) and (\ref{extrafive}) 
over Eqs.~(\ref{extraeight}) and (\ref{extranine}), as CMNT appear to suggest, 
then the difference in the 
resulting cross sections can be interpreted as a measure of theoretical 
uncertainty.  This interpretation would not justify firm 
conclusions of a minimal $0.8\%$ increment in the physical cross section 
based on the choice of Eqs.~(\ref{extrafour}) and (\ref{extrafive}).   

%As remarked at the beginning of this section, 
The CMNT value for the inclusive top quark cross 
section at $m=175\ {\rm GeV}$ and $\sqrt{S}=1.8\ {\rm TeV}$, including 
theoretical uncertainty, lies within our uncertainty band.  Therefore, the 
numerical differences between our results for top quark production 
at the Tevatron have little practical significance.  However, there are 
important differences of principle in our treatment of subleading 
contributions that will have more significant consequences for predictions in 
other processes or at other values of top mass and/or at other energies, 
particularly in reactions dominated by $gg$ subprocesses.  

\section{Discussion and Conclusions}

The advantages of the perturbative resummation method~\cite{previous} are that 
there are no arbitrary infrared cutoffs and there is a well-defined 
region of applicability where subleading logarithmic terms are suppressed. 
When evaluated for top quark production at ${\sqrt S}=1.8$ TeV,  
our resummed cross sections are about $9\%$ above the next-to-leading order
cross sections computed with the same parton distributions. The 
renormalization/factorization scale 
dependence of our cross section is fairly flat, resulting in a $9-10\%$ 
theoretical uncertainty.  
%This variation is smaller than the corresponding 
%dependence of the next-to-leading cross section, as should be expected.
Our perturbative boundary of 1.22 GeV above the threshold in the dominant 
$q\bar q$ channel is comparable to the hadronic width of the top quark, a 
natural definition of the perturbative boundary. 
% In recent 
%papers\cite{catani}, the authors state that the increase in cross section 
%they find with their resummation method is no more than $1\%$ over 
%next-to-leading order.  The numerical difference in the two approaches 
%boils down to the treatment of the subleading logarithms, which can 
%easily shift the results by a few percent, if proper care is not
%taken. Our approach includes the universal leading
%logarithms only while theirs includes non-universal subleading
%structures which produce the suppression they find. 

%Our theoretical analysis and the stability of our cross sections under $\mu$
%variation provide confidence that our perturbative resummation procedure 
%yields an accurate calculation of the inclusive top quark cross section at 
%Tevatron energies and exhausts present understanding of the perturbative 
%content of the theory.  
Our estimated theoretical 
uncertainty of $9-10\%$ is associated with $\mu$ variation.  An entirely 
different procedure to estimate the overall theoretical uncertainty is to 
compare our enhancement of the cross section above the next-to-leading order 
value to that of CMNT~\cite{catani}, again yielding about $10\%$.  An 
interesting question is whether theory can aspire to an accuracy of better 
than $10\%$ for the calculation of the top quark cross section.  To this end, 
a mastery of subleading logarithms would be desirable, perhaps 
requiring a formidable complete calculation at next-to-next-to-leading order of 
heavy quark production, to establish the possible pattern of subleading 
logarithms, and resummation of both leading and subleading 
logarithms.  An analysis in moment space of the issues involved in resummation 
of next-to-leading logarithms for heavy quark production is presented by 
Kidonakis and Sterman~\cite{kidster}.  Inversion of the resummed moments to 
the physically relevant momentum space requires considerable work.
Full implementation of the resummation of next-to-leading 
logarithms would reduce the difference somewhat between our results and 
those of CMNT and move the debate to the level of 
next-to-next-to-leading logarithms.  

Our prediction falls within the relatively large experimental uncertainties.  
%Despite the different treatment of subleading 
%terms, our calculation of the inclusive cross section for top quark production 
%at the Fermilab Tevatron and that of Ref.~\cite{catani} fall within the 
%estimated uncertainties of each other.  
If a cross section significantly 
different from ours is measured in future experiments at the Tevatron with 
greater statistical precision, we would look for explanations in effects  
beyond QCD perturbation theory.  These explanations might include unexpectedly 
substantial non-perturbative effects or new production mechanisms.  An 
examination of the distribution in $\eta$ might be revealing.

The all-orders summation of large logarithmic terms, that are 
important in the near-threshold region of small values of the scaled partonic 
subenergy, $\eta \rightarrow 0$, was described here for the specific case of 
top quark production at the Fermilab Tevatron collider.  Other processes for 
which threshold resummation will also be pertinent include
the production of hadronic jets that carry large values of transverse momentum 
and the production of pairs of supersymmetric particles with large mass.  

\section*{Acknowledgments}

I am most appreciative of the warm hospitality extended by Professor Jiro 
Kodaira and his colleagues at Hiroshima University.  
The research described in this paper was carried out in collaboration with 
Dr. Harry Contopanagos whose current address is 
Electrical Engineering Department, University of California  
at Los Angeles, Los Angeles, CA 90024.  
Work in the High Energy Physics Division at Argonne National Laboratory is 
supported by the U.S. Department of Energy, Division of High Energy Physics, 
Contract W-31-109-ENG-38. 
%\vspace{-0.2cm}

\section*{References}
%\vspace{-0.2cm}

\end{document}